\documentclass[twocolumn]{aastex61}

\newcommand\aastex{AAS\TeX}

\usepackage{enumitem}
\usepackage{graphicx}

\submitjournal{ApJ}

\shorttitle{\aastex\ Recurrent Jetlets Associated with the Disappearance of a Satellite Spot}
\shortauthors{Yang et al.}

\begin{document}

\title{Recurrent Jetlets Associated with the Disappearance of a Satellite Spot}

\correspondingauthor{Liheng Yang}
\email{yangliheng@ynao.ac.cn}

\author[0000-0002-0786-7307]{Liheng Yang}
\affil{Yunnan Observatories, Chinese Academy of Sciences, Kunming 650216, China}
\affiliation{Yunnan Key Laboratory of the Solar physics and Space Science, Kunming 650216}

\author{Xiaoli Yan}
\affiliation{Yunnan Observatories, Chinese Academy of Sciences, Kunming 650216, China}
\affiliation{Yunnan Key Laboratory of the Solar physics and Space Science, Kunming 650216}

\author{Jun Zhang}
\affiliation{School of Physics and Optoelectronics Engineering, Anhui University, Hefei 230601, China}

\author{Zhike Xue}
\affiliation{Yunnan Observatories, Chinese Academy of Sciences, Kunming 650216, China}
\affiliation{Yunnan Key Laboratory of the Solar physics and Space Science, Kunming 650216}

\author{Zhe Xu}
\affiliation{Yunnan Observatories, Chinese Academy of Sciences, Kunming 650216, China}
\affiliation{Yunnan Key Laboratory of the Solar physics and Space Science, Kunming 650216}

\author{Jincheng Wang}
\affiliation{Yunnan Observatories, Chinese Academy of Sciences, Kunming 650216, China}
\affiliation{Yunnan Key Laboratory of the Solar physics and Space Science, Kunming 650216}

\author{Yijun Hou}
\affiliation{National Astronomical Observatories, Chinese Academy of Sciences, Beijing 100101, China}

\author{Yian Zhou}
\affiliation{Yunnan Observatories, Chinese Academy of Sciences, Kunming 650216, China}
\affiliation{Yunnan Key Laboratory of the Solar physics and Space Science, Kunming 650216}

\author{Defang Kong}
\affiliation{Yunnan Observatories, Chinese Academy of Sciences, Kunming 650216, China}
\affiliation{Yunnan Key Laboratory of the Solar physics and Space Science, Kunming 650216}

\author{Roslan Umar}
\affiliation{East Coast Environmental Research Institute (ESERI), Universiti Sultan Zainal Abidin, 21300 Kuala Nerus,Terengganu, Malaysia}

\author{Xinsheng Zhang}
\affiliation{Yunnan Observatories, Chinese Academy of Sciences, Kunming 650216, China}
\affiliation{Yunnan Key Laboratory of the Solar physics and Space Science, Kunming 650216}

\author{Qiaoling Li}
\affiliation{Department of Physics, Yunnan University, Kunming, Yunnan 650091, China}

\author{Liping Yang}
\affiliation{School of Physics, Electrical and Energy Engineering, Chuxiong Normal University, Chuxiong 675000, China}

\begin{abstract}

Recurrent small-scale eruptions are fascinating phenomena in the solar atmosphere, characterized by repeated energy buildup and release over short time intervals. However, their underlying physical mechanisms remain unclear. On 2021 May 23, five recurrent jetlets (J1-J5) were observed continuously ejecting from a satellite spot located at the north edge of AR 12824. Using high-resolution, multi-wavelength data from NVST, SDO, and IRIS, we investigate the physical characteristics of these jetlets and their relationship with the satellite spot. The widths of these jetlets range from 1300 to 2900 km, their lifetimes range span 3 to 10 minutes, and their projection speeds vary from 152.8 to 406.0 km s$^{-1}$. During the eruptions, the satellite spot moved northwest at a low speed of 376 $\pm$ 12 m s$^{-1}$. Its area gradually decreased due to magnetic cancellation with surrounding positive magnetic field, resulting in an average cancellation rate of 1.3$\times$10$^{18}$ Mx hr$^{-1}$. Dark lanes that separated from the satellite spot and small pores were observed to move toward nearby these features or dark lanes with opposite polarities, eventually disappearing during the magnetic cancellation process. J4 was driven by an eruption of a micro-filament. Spectral observations revealed a redshift on the right side of J4 and a blueshift on the left side of its base, suggesting a counterclockwise rotation. The horizontal magnetic field of the satellite spot consistently exhibited a vortex structure throughout its evolution until it vanished. The nonlinear force-free field extrapolation confirms that the satellite spot serves as one footpoint of a mini-flux rope. These observations reveal that these jetlets might result from three-dimensional null-point magnetic reconnection, initiated by the continuous eruption of a mini-flux-rope or multiple mini-flux-ropes, driven by sustained magnetic cancellation.

\end{abstract}


\section{Introduction}

Jets are small-scale explosive phenomena on the Sun, representing a transient energy release process. Their observation dates back to the 1930s \citep{1934MNRAS..94..472N}. Since the launch of Yohkoh in 1991, jets have been extensively studied in observations and theory \citep[e.g.,][]{1992PASJ...44L.173S,1995Natur.375...42Y,1996PASJ...48..123S}. With the improvement in observational resolution, more fine structures of jets have been detected, such as plasma blobs \citep[e.g.,][]{2014A&A...567A..11Z,2019ApJ...885L..15K,2019ApJ...870..113Z,2022FrASS...8..238C}, current sheets \citep[e.g.,][]{2018ApJ...854..155K,2019ApJ...879...74C,2023ApJ...942...86Y,2024MNRAS.528.1094Y}, Kelvin–Helmholtz instability \citep[e.g.,][]{2018NatSR...8.8136L}, and cool and hot components\citep[e.g.,][]{2007A&A...469..331J,2017ApJ...851...67S}. Jets are prevalent in coronal holes, quiet Sun regions, and active regions, where they might provide mass and energy to upper corona \citep[e.g.,][]{2014Sci...346A.315T}. As a result, they are considered potential candidates for coronal heating and solar wind acceleration \citep[e.g.,][]{1983ApJ...272..329B}. Sometimes, their generation might be the result of magnetic reconnection between mini-filaments and large-scale active region loops, filaments, or filament channel \citep[e.g.,][]{2019ApJ...887..220Y,2019ApJ...887..239Y,2024ApJ...964....7Y}. As analogues of large-scale eruptions, coronal jets can also excite coronal mass ejections and extreme ultraviolet waves \citep[e.g.,][]{2021ApJ...911...33C,2012ApJ...745..164S,2022ApJ...931..162Z,2022ApJ...926L..39D,2024ApJ...968..110D}. High-resolution observations have revealed increasing smaller-scale jets, such as network jets \citep[e.g.,][]{2014Sci...346A.315T}, jetlets \citep[e.g.,][]{2014ApJ...787..118R, 2019ApJ...887L...8P}, nanojets \citep[e.g.,][]{2021NatAs...5...54A}, picoflare jets \citep[e.g.,][]{2023Sci...381..867C}, and references therein. The newly discovered picoflare jets in coronal holes are believed to produce enough high-temperature plasma to sustain the solar wind. The physical properties, driving mechanisms, and theoretical models of the jets have been summarized in detail by several researches \citep{2016AN....337.1024I,2016SSRv..201....1R,2021RSPSA.47700217S,2022AdSpR..70.1580S}.

Jetlets are small-scale UV/EUV ejections that are intermediate in size between coronal X-ray jets \citep[$\sim$8000 km in width;][]{2007PASJ...59S.771S} and chromospheric spicules \citep[$\sim$200 km in width;][]{2007PASJ...59S.655D}. They were first detected and named by \citet{2014ApJ...787..118R} during their study of the role of transients in the sustainability of coronal hole plumes. In their research, jetlets were observed at the footpoints of plumes, with durations ranging from tens of seconds to a few minutes. The discovery of these jetlets led to the hypothesis that they might be the long-awaited source of mass and energy that sustains plumes over periods of hours to weeks. Further study by \citet{2018ApJ...868L..27P} revealed that jetlets are not confined to the base of plumes but also widespread in the network region. They proposed that jetlets are scaled-down versions of coronal jets, as they shared many similarities with coronal jets, such as similar trigger mechanisms, base brightenings, and rotation motions. The jetlets they investigated had average lengths of 27000 km, spire widths of 3000 km, lifetimes of 3 minutes, and speeds of 70 km s$^{-1}$. Additionally, the high-resolution Coronal Imager 2.1 (Hi-C 2.1) detected smaller jetlets with average lengths of 9000 km, widths of 600 km, and speeds of 60 km s$^{-1}$ \citep{2019ApJ...887L...8P}. Jetlets are known to drive propagating disturbances (PDs) within plumes, which are often interpreted as slow magnetoacoustic waves \citep{2015ApJ...807...71P}. These PDs occur in phase with the jetlet brightenings, which typically show a periodicity of 3–5 minutes, closely matching the oscillations observed in photospheric and chromospheric p-modes \citep{2022ApJ...933...21K}. Recently, \citet{2023ApJ...945...28R} presented evidence that widespread jetlets, driven by small-scale magnetic reconnection near the base of the corona, could play a crucial role in heating and accelerating the solar wind.

Coronal jets are usually caused by miniaturized versions of filament eruptions, referred to as mini-filaments \citep[e.g.,][]{2019ApJ...874..146H, 2019ApJ...881..132D}. These mini-filaments typically have a size of 19000 km and a lifetime of 50 minutes \citep{2000ApJ...530.1071W}. \citet{2010ApJ...720..757M} classified the coronal jets resulting from the mini-filaments as blowout jets. Another type is the classical coronal jets, which are formed through the magnetic reconnection between the newly emerging closed magnetic field and surrounding open magnetic field \citep{1992PASJ...44L.173S,1995Natur.375...42Y}. \citet{2015Natur.523..437S} point out that all the coronal jets are triggered by mini-filament eruptions. Successful eruptions lead to blowout jets, while failed eruptions result in classical coronal jets. In a classical coronal jet, the core field of the base arch lacks any twist. The jet's spire and bright point are typically located at opposite ends of the base arch, 
activating and expanding in synchrony. Their lateral extent is significantly smaller than that of the base arch. As the jet develops, no discernible heating of X-ray plasma is observed within the interior of the base arch \citep{2010ApJ...720..757M}. As analogues of coronal jets, jetlets are expected to be associated with the eruption of a small filament. However, only one observed case of an erupting small filament has been found to produce a jetlet \citep{2022ApJ...933...21K}. In this instance, the analyzed jetlet was twice the size of a typical jetlet, but still smaller than typical coronal jets. Therefore, it remains uncertain whether jetlets are truly driven by small-scale filaments. \citet{2016ApJ...828L...9S} investigated the number–size relationship of solar filament-eruption-like events, including those as small as hypothesized microfilament eruptions responsible for spicule formation. Their analysis revealed a power-law distribution, lending credence to the theory that spicules may stem from such micro-scale eruptions. More recently, \citet{2024ApJ...963....4S} included additional number-versus-size data for jetlets, finding it consistent with the previously established distribution when accounting for error bars. This suggests that jetlets are likely driven by small-scale filament eruptions, with the estimated scale of such filaments ranging from 500 km to 5000 km.

Magnetic reconnection plays a crucial role in jet eruptions across various scales, as it involves the rearrangement of magnetic field lines and the conversion of magnetic energy into kinetic and thermal energy. However, 
the exact triggering and driving mechanism remain not fully understood. One possible mechanism is magnetic flux emergence, which introduces new magnetic energy into the solar atmosphere, triggering 
interchange reconnection with the surrounding magnetic fields and generating jets \citep{1992PASJ...44L.173S,1995Natur.375...42Y,2008ApJ...673L.211M}. Several jets have been observed to occur simultaneously with 
episodes of flux emergence \citep[e.g.,][]{1996ApJ...464.1016C,2007A&A...469..331J}. Additionally, a growing body of observations supports flux cancellation as another potential trigger for reconnection that drives coronal jets \citep[e.g.,][]{2011ApJ...738L..20H,2012NewA...17..732Y,2016ApJ...832L...7P,2018ApJ...853..189P,2018ApJ...852...10L,2019ApJ...882...16M,2021ApJ...919...34Y}. Magnetic flux cancellation occurs when opposite-polarity magnetic 
fragments on the photosphere converge and submerge, often leading to magnetic reconnection in the overlying atmosphere, which can subsequently trigger coronal jets \citep[e.g.,][]{2024ApJ...960...51P}.
For jetlets, nearly all studies provide observational evidence that magnetic cancellation contributes to jetlet formation \citep{2014ApJ...787..118R,2018ApJ...868L..27P,2019ApJ...887L...8P,2023ApJ...945...28R}. However, due to the limited 
number of cases, further confirmation is needed.

Various models have been proposed to explain the formation of coronal jets. One widely accepted model, the emerging-flux model, posits that coronal jets are generated through magnetic reconnection between emerging magnetic bipoles 
and the surrounding oblique open field \citep[e.g.,][]{1992PASJ...44L.173S,1995Natur.375...42Y,2005ApJ...635.1299A,2008ApJ...673L.211M}. \citet{2009ApJ...691...61P} introduced the embedded-dipole model, in which a localized 
region with strong magnetic polarity embedded within an ambient field of opposite polarity, is gradually rotated, storing magnetic free energy and helicity in the corona. Ultimately, a kink-like instability triggers explosive interchange reconnection 
between the twisted closed field and the ambient untwisted open field, leading to the formation of a twisted jet. Continued stress applied at the photospheric boundary can then generate recurrent, untwisting quasi-homologous jets \citep{2010ApJ...714.1762P}. 
More recently, \citet{2017Natur.544..452W} proposed a breakout model for coronal jets. In this model, a filament channel forms beneath the spine-fan structure of a three-dimentional null point, and breakout reconnection occurring at the null triggers the eruption of the filament channel, resulting in the formation of a twisted jet. Subsequent work revealed that the twisted jets observed at the peripheries of active regions arise from a combination of magnetic breakout and ideal kink instability of the erupting flux rope \citep{2019MNRAS.490.3679W}. Additionally, some studies have extended these models into interplanetary space by adopting spherical geometry, investigating the potential interplanetary effects caused by solar jets \citep[e.g.,][]{2017ApJ...834...62K, 2017ApJ...834..123S}. As analogues of coronal jets, jetlets are expected to share similar formation mechanisims.

In this paper, five recurrent jetlets ejecting from the north edge of the active region NOAA 12824 have been investigated in detail with high-resolution, multi-wavelength observations from the New Vacuum Solar Telescope \citep[NVST;][]{2014RAA....14..705L,2020ScChE..63.1656Y}, the Solar Dynamics Observatory \citep[SDO;][]{2012SoPh..275....3P}, and the Interface Region Imaging Spectrograph \citep[IRIS;][]{2014SoPh..289.2733D}. Their physical characteristics, potential relationships with magnetic cancellation, as well as the evolution detail of a satellite spot during the magnetic cancellation, are unveiled. This paper is organized as follows. Section 2 provides a brief overview of the observational data. The findings are presented in Section 3, followed by the conclusions and discussions in Section 4.

\section{Observations and Data Analysis}
 
The IRIS contains a 19 cm ultra-violet (UV) telescope that can provide simultaneous spectra and images from photosphere up to corona with a spatial resolution of 0.33 $''$-0.4 $''$. It observes slit-jaw images in four passbands with one photospheric passband (2830~{\AA}), one chromospheric line (Mg $\tiny{\uppercase\expandafter{\romannumeral2}}$ k 2796~{\AA}), and two transition-region lines (C $\tiny{\uppercase\expandafter{\romannumeral2}}$ 1335~{\AA} and Si $\tiny{\uppercase\expandafter{\romannumeral4}}$ 1400~{\AA}). It can obtain spectral in far ultra-violet (FUV) passbands from 1332~{\AA} to 1358~{\AA} and in near ultra-violet (NUV) passbands from 2783~{\AA} to 2835~{\AA}, covering a temperature diagnostics from 5000 K to 10 MK. On 2021 May 23, the IRIS targeted the active region NOAA 12824 during 02:29-03:18 UT. It conducted a very large dense 320-step raster with a step cadence of 9.2 s and a raster cadence of 2.937 s. The width of the IRIS slit is 0.35 $''$. The slit-draw images are obtained simultaneously with the spectra in four wavelengths (1330~{\AA}, 1400~{\AA}, 2796~{\AA}, and 2832~{\AA}) with a cadence of 37 s. 

The NVST is a one-meter vacuum telescope located near Fuxian Lake in Southwest China. It is designed to observe the fine structures of the Sun's photosphere and chromosphere. The telescope consists of a multi-channel high resolution imaging system, a multi-band spectrometer and an adaptive optics (AO) system. The multi-channel high resolution imaging system is mounted on a 6-m rotating platform. It observes the photosphere in the TiO line at 7058~{\AA}, and the chromosphere in the H$\alpha$ and He I lines at 6563~{\AA} and 10830~{\AA}. A tunable Lyot filter is placed in the H$\alpha$ optical path, allowing spectral scans within a range of $\pm$5~{\AA} with a step size of 0.1~{\AA}, and a bandwidth of 0.25~{\AA}. The multi-band spectrometer is installed on the vertical hanging bracket right below the rotating platform in two passbands (Fe I 5324~{\AA} and Ca II 8542~{\AA}). The AO system is positioned before the other two instruments, and is a high order system with 151 actuators \citep{2016ApJ...833..210R,2016SPIE.9909E..2CZ,2023SCPMA..6669611Z}. The Strehl ratio of the corrected image exceeds 0.75 when seeing is better than or equal to 10 cm. On 2021 May 23, the NVST focused on the active region NOAA 12824 during 02:00-03:59 UT. Five recurrent jetlets were detected in the H$\alpha$ 6563~{\AA} data. These H$\alpha$ images have a pixel size of about 0.165 $''$ and a cadence of 12 s. The data were processed by subtracting the dark current, correcting for flat field, and then reconstructed using the speckle masking method \citep{2016NewA...49....8X,2022RAA....22f5010C}. TiO data were also used to study the evolution of the satellite spot associated with the recurrent jets. These data have a pixel size of about 0.052 $''$ and a cadence of 30 s. The NVST TiO images were converted to heliographic coordinates by registering them to the Helioseismic and Magnetic Imager (HMI) continuum image using an automatic mapping technique \citep{ji19}. Photospheric flow field data were obtained using the dense optical flow method from the OpenCV library (cv2.calcOpticalFlowFarneback), based on Gunnar Farneback's algorithm \citep{gun2003}. The averaging window size was set to 11 pixels, corresponding to a spatial scale of about 0.6 $''$. 

The SDO consists of three instruments, including the Atmospheric Imaging Assembly \citep[AIA;][]{2012SoPh..275...17L} , Extreme Ultraviolet Variability Experiment \citep[EVE;][]{2012SoPh..275..115W}, and HMI \citep{2012SoPh..275..207S}. The AIA can provide full-disk images of seven extreme ultraviolet (EUV) wavebands (94~{\AA}, 131~{\AA}, 335~{\AA}, 211~{\AA}, 193~{\AA}, 171~{\AA}, and 304~{\AA}) and three continuum bands (4500~{\AA}, 1700~{\AA}, and 1600~{\AA}), covering a temperature diagnostics from 0.005 to 20 MK. The HMI provides full-disk Doppler velocity, line-of-sight magnetic flux, and continuum proxy images with 1$''$ resolution every 45 seconds, as well as vector magnetic field maps every 12 minutes. Using a nonlinear force-free field (NLFFF) method, the three-dimensional magnetic field of the Sun's atmosphere is reconstructed by extrapolating vector magnetic field data \citep{2004SoPh..219...87W,2006SoPh..233..215W}. The Doppler velocity images are corrected by removing SDO motion and solar rotation \citep{2013ApJ...765...98W}. 

\section{Results}

\subsection{The Source Region of the Recurrent Jetlets}

On 2021 May 23, five recurrent jetlets were observed to continuously eject from a satellite spot on the north edge of the active region NOAA 12824. Satellite spots are small spots with opposite polarity that 
appear within or near the penumbra of a main sunspot \citep[e.g.,][]{1996ApJ...464.1016C,2015ApJ...815...71C}. This active region is located at N21$^\circ$W01$^\circ$, which is classified as a $\beta$$\gamma$-type configuration. 
Figure 1 illustrates the positions of these jetlets along with their adjacent magnetic environment. Their source region is circled by a purple square. It is clear from the magnetogram that the source region is situated at the edge of the 
positive magnetic network (marked by red arrows in Figure 1(a)). It is noted that there is one satellite spot with negative polarity in the source region (marked by green arrows in 
Figures 1 (a) and (b)). The negative polarity, being encircled by positive polarity, implies a fan-spine magnetic configuration. The generation of the jetlets are associated with the movement of this satellite spot. 
The first jetlet is displayed in the NVST H$\alpha$ (Figure 1(c)) and SDO/AIA 193~{\AA} images (Figure 1(d)). It appears to occur beneath the large-scale active region loops (see Figure 1(d)). In H$\alpha$ wavelength, these 
large-scale active region loops manifest as dark long fibrils (Figure 1(c)). The jetlet shows up as a bright long, thin structure in the 193~{\AA} wavelength, but a dark structure in H$\alpha$ wavelength, suggesting that the jetlet plasma 
is heated during its eruption. At the base of the jetlet, a brightening is observed (see Figure 1(d)) , similar to the base bright points of coronal jets.

\subsection{The Physical Characteristics of the Recurrent Jetlets}

Figure 2 presents the five recurrent jetlets (marked by white arrows) in H$\alpha$, 1400~{\AA}, 1330~{\AA}, 304~{\AA} and 131~{\AA} wavelengths, respectively. These jetlets are marked with the letter ``J'' followed by the jetlet number. 
It is important to highlight that only the final four jetlets were detected by IRIS. Notably, the five jetlets almost have the same shapes and eruptive trajectories, but different sizes (see Animation 1). J4 seems to be the strongest 
eruption. As mentioned above, the generation of the jetlets is related to the movement of the satellite spot. Consequently, the source region of the jetlets is not fixed. The contours of the satellite spot
are superimposed on the 131~{\AA} images, as shown in Figures 2(d1) -- (d5). It is evident that the satellite spot moved towards the northwest, and concurrently, the base brightenings also underwent a gradual shift in the same northwest 
direction as the satellite spot progressed. The displacement in both the west and north directions is within 2$''$. Interestingly, the jetlets in 1400~{\AA} and 1330~{\AA} 
wavelengths are not sheltered by large-scale active region loops. They exhibit a clear inverted-Y shape (marked by green arrows), characteristic of typical coronal jets and widely recognized as an indicator of 
magnetic reconnection \citep{2007Sci...318.1591S}.

To investigate the kinematics characteristics of these jetlets, we created time-distance diagrams (shown in Figure 3) using H$\alpha$, 1400~{\AA}, 304~{\AA}, 171~{\AA}, 193~{\AA}, 335~{\AA}, 94~{\AA}, and 131~{\AA} images 
along the propagation path of the jetlets, indicated by the red line in Figure 1(c). As observed in these diagrams, the jetlets appear as dark strips in the H$\alpha$ wavelength, while they show bright strips in the other hotter bands, 
which cover a temperature range from approximately 0.1 MK to 20 MK. Two points were chosen along each strip to calculate the jetlet projected speed. We conducted the measurements 10 times, determined the average as 
the final speed, and used the standard deviation as the error. It can be seen that all the jetlets exhibit similar speeds across different wavelengths. J3 has the highest speed, ranging from 406.0$\pm$37.9 km s$^{-1}$ to 
433.2$\pm$28.6 km s$^{-1}$, while J1 has the lowest speed, from 146.6$\pm$11.2 km s$^{-1}$ to 158.6$\pm$10.1 km s$^{-1}$. J5 was not detected in the 94~{\AA} wavelength. However, its speed was estimated to be 
between 152.8$\pm$6.4 km s$^{-1}$ and 164.3$\pm$10.3 km s$^{-1}$ in other wavelength channels. It exhibits a noticeable deceleration in H$\alpha$ and other AIA wavelengths, and its speed decreased to about half of its original value, ranging 
from 60.7$\pm$3.9 km s$^{-1}$ to 66.6$\pm$8.8 km s$^{-1}$. J2 was detected in all eight wavelengths, showing a speed range from 208.0$\pm$18.9 km s$^{-1}$ to 242.9$\pm$16.2 km s$^{-1}$. 
The plasma of J3 and J4 was intensely heated, making it difficult to trace their trajectories in H$\alpha$ wavelength. J4 has a speed range of 406.0$\pm$37.9 km s$^{-1}$ to 425.5$\pm$30.6 km s$^{-1}$ in the other seven 
wavelengths. Unlike some observations of coronal jets, no falling jetlet material was observed.

Table 1 lists the physical parameters of the five jetlets, including the jetlet number, start time, lifetime, speed, spire width, and jetlet base width. It can be seen that four of the jetlets have a lifetime of 3 minutes, while only one 
lasts a longer time, about 10 minutes, which is consistent with the typical jetlet lifetimes \citep{2023ApJ...945...28R}. The jetlet spire width is relatively narrow, ranging from 1300 km to 2900 km, which is much smaller than the typical 
coronal jet width \citep[Tens of thousands to hundreds of thousands of km;][]{2023ApJ...945...96Y}. Additionally, we note that the jet base width is comparable to the jet spire width, ranging from 1300 km to 2600 km. 
The speed of the jetlets ranges from 152.8 km s$^{-1}$ to 406.0 km s$^{-1}$, which is much higher than the speed of typical jetlets \citep[150 km s$^{-1}$;][]{2023ApJ...945...28R}. 
 Based on the simulations by \citet{2009ApJ...691...61P}, the acceleration of jet plasma is primarily driven by nonlinear torsional Alfv$\acute{e}$n waves. In this framework, the maximum plasma velocity is approximately 
 equal to the local Alfv$\acute{e}$n speed, which depends on both the magnetic field strength and plasma density. Therefore, the variation in jetlet speeds is likely a result of differences in these local physical parameters.

\subsection{An example of jetlet eruptions}

\subsubsection{The Evolutions of J4}

Among all the jetlets, J4 is the widest and has the longest duration. It is also observed by IRIS, making it an ideal candidate for detailed study, as shown in Figure 4. Around 02:58:14 UT, a mini-filament-like structure was observed rising at the 1400~{\AA} wavelength (indicated by a blue arrow in Figure 4(b1); see Animation 2). However, due to the influence of the previous jetlet, its exact position before eruption could not be determined. Its size is estimated to be comparable to the base of J4 at the 1400~{\AA} wavelength, approximately 6000 km, which is significantly smaller than the typical length of mini-filaments \citep[about 19000 km;][]{2000ApJ...530.1071W}. The micro-filament was not detected in the H$\alpha$ and 304~{\AA} wavelengths, likely due to being obscured by dense overlying fibril structures. Almost simultaneously, a brightening appeared at the base of the jetlet in the 94~{\AA} wavelength (marked by the red arrow in Figure 4(d1), see Animation 2). One minute later, a thin, bright spire 
launched from the base of the jetlet. This spire appeared as a bright, slightly curved structure clearly visible across all four wavelengths (marked by white arrows in panels (a2)-(d2); see animation 2), indicating that it is a multi-thermal structure. 
The base brightening was circular in shape (marked by sky blue arrows in Figures 4(c2) and (d2)), and gradually expanded over time (see Figures 4(a2)–(d2)). In the 1400~{\AA} channel, the jetlet displayed an inverted-Y shape (marked by white arrows in panels (b2)-(b4)), resembling coronal jets. As noted earlier, the magnetic configuration in the source region of the jetlet is favorable for the development of a fan-spine topology. The presence of both the circular base brightening and the inverted-Y shape 
suggests that three-dimensional null-point magnetic reconnection may have occurred, consistent with previous simulation results of \citet{2009ApJ...691...61P} and \citet{2017Natur.544..452W}. Around 03:03:05 UT, a dark filamentary structure appeared within the jetlet (marked by the blue arrows in Figures 4(c3) and (c4)), suggesting the injection of cool material. However, as noted, the dark features in Figures 4(c3) and (c4) do not provide conclusive evidence. Estimating the temperature of such cool material remains challenging due to the limitations of the observations. While Differential Emission Measure (DEM) analysis is a standard method for inferring the thermal structure of emitting plasma, it is not applicable to absorbing material. Cool plasma typically manifests as absorption features (i.e., dark structures) against the EUV background and does not emit strongly in these wavelengths, making it difficult to constrain its thermal properties using DEM. Therefore, we are unable to provide a quantitative temperature estimate for the cool material in this instance. This cool material might originate from the micro-filament and is also visible in the H$\alpha$ wavelength (marked by a red arrow in Figure 4(a4)). The jetlet appeared to exhibit a counterclockwise rotation (see the H$\alpha$ movie in Animation 2). Over time, the jetlet traveled further along the outer spine of the fan-spine topology. Five minutes later, the jetlet had completely dissipated.

\subsubsection{The Spectrum Features of J4}

The IRIS slit scanned J4 between 03:02 UT and 03:07 UT. We selected the 03:04 UT spectra for detailed analysis because the slit simultaneously captured the jetlet and its base. The IRIS slit is nearly parallel to the jetlet axis, as indicated by the black dashed line on the 1400~{\AA} slit-draw image in Figure 5. Note that the jetlet (marked by a green line in Figures 5(b1)-(d1)) exhibits a prominent redshift in the transition region line (Si $\tiny{\uppercase\expandafter{\romannumeral4}}$ 1402~{\AA}), but only a weak redshift in the low transition region line (C $\tiny{\uppercase\expandafter{\romannumeral2}}$ 1335~{\AA}) and the chromospheric line (Mg $\tiny{\uppercase\expandafter{\romannumeral2}}$ k 2796~{\AA}). In contrast, the jetlet base (marked by a blue line in Figures 5(b1)-(d1)) shows a clear blueshift in all three spectral lines. Two positions, marked with a green and a blue pluses in Figure 5(a), are chosen to study the observed spectral profiles. The profiles of the Si $\tiny{\uppercase\expandafter{\romannumeral4}}$ line clearly exhibit a Gaussian shape. Therefore, we fit the two profiles with single-Gaussian function. The line center is determined using a single Gaussian fit applied to the averaged spectrum along the white vertical bar in panel (a), and is calculated to be 1402.7723~{\AA}. The fitting results are overplotted with dashed and solid black lines in Figures 5(b2)-(b3). The Doppler velocity and the line width of the jetlet are 25.2 km \textbf{s$^{-1}$} and 0.57~{\AA}, respectively. For the jetlet base, the Doppler velocity and the line width are -35.6 km \textbf{s$^{-1}$} and 0.53~{\AA}, respectively. The C $\tiny{\uppercase\expandafter{\romannumeral2}}$ 1335~{\AA} and the Mg $\tiny{\uppercase\expandafter{\romannumeral2}}$ k 2796~{\AA} lines are optically thick and exhibit self-reversed line profiles (see Figures 5(c2), (c3), (d2), and (d3)). As a result, the physical parameters (Doppler velocity and line width) are estimated using the center-of-gravity method. 
The reference line centers are determined by applying this method to the averaged spectra along the white vertical bar in panel (a). The reference line centers of the two lines are 1335.7367~{\AA} and 2796.2835~{\AA}, respectively. 
The Doppler velocity and width of the jetlet are 4.3 km \textbf{s$^{-1}$} and 0.43~{\AA} in C $\tiny{\uppercase\expandafter{\romannumeral2}}$ 1335~{\AA} line, and 4.1  km \textbf{s$^{-1}$} and 0.51~{\AA} in Mg $\tiny{\uppercase\expandafter{\romannumeral2}}$ k 2796~{\AA} line. The jetlet base has a Doppler velocity of -35.9 km \textbf{s$^{-1}$} and a width of 0.53~{\AA} in C $\tiny{\uppercase\expandafter{\romannumeral2}}$ 1335~{\AA} line, and a Doppler velocity of -7.9 km \textbf{s$^{-1}$} and a width of 0.55~{\AA} in Mg $\tiny{\uppercase\expandafter{\romannumeral2}}$ k 2796~{\AA} line. The redshift of the jetlet and the blueshift of its base appear to indicate a counterclockwise rotation of the jetlet. The helicity sign of the jetlet should be negative with left-handed chirality. This rotational motion seems to be more prominent in the transition region.

The 03:04 UT spectra also contain the O $\tiny{\uppercase\expandafter{\romannumeral4}}$ density-sensitive pair of lines at 1399.77~{\AA} and 1401.16~{\AA}, as shown in Figure 6(a). Using the CHIANTI atomic database v8.0, we determine the electron number density based on the ratio of the intensities of these two O $\tiny{\uppercase\expandafter{\romannumeral4}}$ lines \citep{2015A&A...582A..56D}. The theoretical intensity ratio (1399.77~{\AA}/1401.16~{\AA}) as a function of electron number density is displayed in Figure 6(b). Since the signals of both the jetlet and its base in the O $\tiny{\uppercase\expandafter{\romannumeral4}}$ 1399.77 Å line are weak, we use the average spectrum over a specific range to calculate the density. The jetlet profile (green line), averaged between the two green lines in Figure 6(a), along with its corresponding double-Gaussian fit (black line), is shown in Figure 6(c). The intensity ratio (1399.77~{\AA}/1401.16~{\AA}) is estimated to be approximately 0.269, corresponding to a density of 4.78 $\times$ $10^{10}$ $cm^{-3}$ (indicated by the green dotted lines in Figure 6(b)). The profile of jetlet base (blue line), and its corresponding double-Gaussian fit (black line) are presented in Figure 6(d). The intensity ratio of the two O $\tiny{\uppercase\expandafter{\romannumeral4}}$ lines for the jetlet base is calculated to be approximately 0.286, corresponding to a density of 6.34 $\times$ $10^{10}$ $cm^{-3}$ (indicated by the blue lines in Figure 6(b)). The calculated density of J4 and its base is significantly higher than that reported in the previous study by \citet{2022ApJ...933...21K}, likely due to intrinsic physical differences and the application of distinct diagnostic techniques. Our spectral diagnostics yield localized and direct density measurements within compact, high-density regions, whereas their DEM approach averages over larger spatial scales, potentially leading to an underestimation of densities in small-scale structures.

To estimate the kinetic energy of J4, we assume the jetlet is cylindrical. Its width and length are estimate to be 2900 km and 20400 km, respectively, based on the observations. The jetlet's mass is \emph{m} $\approx$ 1.1 $\times$ $10^{13}$ g. Therefore, the kinetic energy is given by K= $\frac{1}{2}$mv$^2$ $\approx$ 5.7 $\times$ $10^{27}$ erg, where v=322.6 km s$^{-1}$. This kinetic energy serves as an upper limit for these jetlets, since J4 is the strongest. The kinetic energy of jetlets surrounding active region is much larger than that in coronal hole, such as 1.3 $\times$ $10^{24}$ erg, as given by \citet{2022ApJ...933...21K}.

\subsection{The photospheric evolution of the source region}

As previously mentioned, these recurrent jetlets are closely associated with a photospheric satellite spot at their base. To further explore this connection, we present the evolution of the satellite spot and its corresponding 
magnetic field at the jetlet base in Figure 7. Prior to the recurrent jetlets, the satellite spot with negative magnetic field (indicated by a blue arrow in Figures 7(a1)-(c1)) and a series of small pores with positive magnetic field were 
observed (see Figure 7(a1) and Animation 3). Over time, the satellite spot gradually moved northwest, while its area steadily decreased. To calculate its velocity, a time-distance diagram was generated along the green line in Figure 7(a1), 
as shown in Figure 7(e). The satellite spot appears as an upward-sloping dark strip in the diagram, with its movement velocity calculated to be 376 $\pm$ 12 m s$^{-1}$. During its motion, magnetic cancellation occurred. 
The negative magnetic field associated with the satellite spot (marked by blue arrows in Figures 7 (a2)-(d2)) decreased continuously, accompanied by a noticeable reduction of the nearby positive magnetic field. To quantify 
the magnetic cancellation, the variations of the negative magnetic flux within the blue box region (see Figure 7(a2)) were calculated, as shown in Figure 7(f). The five dotted lines represent the start time of the five jetlets. 
It is evident that the negative magnetic field experienced a significant decrease during the jetlets, the average flux-loss rate was calculated to be 6.7$\times$10$^{17}$ Mx hr$^{-1}$. Therefore, the average cancellation rate was 
estimated to be 1.3$\times$10$^{18}$ Mx hr$^{-1}$.

Thanks to the high resolution of the TiO images, many detailed evolutionary features of the satellite spot and nearby small pores with positive magnetic field during the magnetic cancellation are revealed, as shown in Figures 8(a1)-(a8). 
It is worth noting that the evolution of dark lanes detaching from these features coincided with the magnetic cancellation (see Animation 4). At 02:00:53 UT, a long dark lane (indicated by a red arrow in Figure 8(a1)) was 
observed moving slowly southeast after separating from the small pores. Seven minutes later, it reached the satellite spot (marked by a red arrow in Figure 8(a2)), where it gradually disintegrated and completely vanished by 
02:17:59 UT (see Figure 8(a3)). During this period, the area of the satellite spot also significantly decreased. At 02:22:30 UT, two dark lanes (marked by sky blue and red arrows in Figure (a4)) were seen separating 
from the satellite spot and small pores. These lanes moved toward each other, collided, and disappeared together around 02:32:03 UT (see Figure 8(a5)). Two minutes later, another dark lane (marked by a sky blue arrow in Figure 8(a6)) separated from the satellite spot and moved toward the small pores, eventually vanishing along with some of them. Following the complete disappearance of the small pores, the remaining satellite spot began to rotate 
and rapidly faded, leaving behind a small area of bright granules indicated by a green arrow in Figure 8(a8); see Animation 4). To further examine the rotational motion, the horizontal velocity field, calculated using the dense optical flow method, is superimposed onto the TiO image at 03:06:42 UT (see Figure 8(a7)), clearly revealing a clockwise rotation of the satellite spot. The averaged horizontal velocity field of the satellite spot is estimated to be approximately 1.96 km s$^{-1}$.

To investigate whether the satellite spot rises or sinks during its evolution, Doppler velocity images are presented in Figures 8(b1)-(b4). The dark contours outline the satellite spot, as depicted in the simultaneous TiO images. It is observed that the satellite spot remained consistently redshifted throughout its evolution, indicating that the satellite spot was sinking. The average Doppler velocity of the satellite spot is from 128.3 m s$^{-1}$ to 139.4 m s$^{-1}$. Figures 8(c1)-(c4) illustrate the evolution of the vector magnetic field associated with the satellite spots. Remarkably, the horizontal magnetic field of the satellite spot consistently exhibited a vortex structure throughout its evolution until it eventually vanished. This suggests a significant distortion in the magnetic field configuration.

Figure 9 shows the NLFFF magnetic topology of the source region from both top (Figures 9(a) and (c)) and side (Figures 9(b) and (d)) views. As seen in Figures 9(a) and (b), a series of twisted magnetic arcades (green lines) are located along the magnetic neutral line of the source region. These arcades likely represent the magnetic structure of a micro-filament, suggesting that non-potential magnetic energy has accumulated in the region. Their average projected length is approximately 5299.6 km, which is comparable to the estimated length of the observed micro-filament ($\sim$ 6000 km). The northern end of the micro-filament is anchored in the parasitic negative polarity region, while the southern end is rooted in the dominant positive polarity region, with its axial field is directed northward. According to the definition provided by \citet{1994ASIC..433..303M}, the micro-filament is dextral. It is expected to possess negative helicity with left-handed chirality \citep{2000ApJ...540L.115C}, which aligns with the helicity sign of J4. As illustrated in Figures 9(c) and (d),  a fan-spine magnetic structure (yellow lines) is located directly above the micro-filament. From the top-down view, the micro-filament appears to be enveloped by the fan-spine structure, which is consistent with observational features. The jetlets observed in this region are likely driven by magnetic reconnection occurring at the three-dimensional magnetic null point associated with the fan-spine topology.

\section{Conclusions and Discussions}

We have presented a detailed analysis of recurrent jetlets originating from the boundary of a network lane surrounding an active region, focusing on their relationship with the evolution of the satellite spot and magnetic cancellation at their base. 
The main findings are summarized as follows.

\begin{enumerate}[label=\arabic*., left=0pt, labelsep=1em, itemsep=0em]

\item These recurrent jetlets, ejected from the edge of an active region, are similar to those originating from coronal holes, as both of them rooted at the network lane. The jetlets studied here had widths of 1300-2900 km, lifetimes of 3-10 minutes, and projection speeds ranging from 152.8 to 406.0 km s$^{-1}$. Their widths and lifetimes fell within the range of typical jetlets reported by \citet{2018ApJ...868L..27P} and \citet{2023ApJ...945...28R}. However, their projected speeds were significantly higher than those of previous jetlets ($\sim$150 km s$^{-1}$). As noted earlier, this variation in jetlet speeds is likely due to differences in magnetic field strength and plasma density. Similar to many coronal jets, these recurrent jetlets exhibit similar shapes and eruptive trajectories. Recurrent jetlets were also observed by \citet{2018ApJ...868L..27P}, who studied jetlets originating from the same polarity inversion lines and exhibiting similar structures.

\item These recurrent jetlets were closely associated with the satellite spot at the jetlet base. During the jetlet eruptions, the satellite spot moved northwest at a low speed of 376 $\pm$ 12 m s$^{-1}$, exhibiting a clear shear motion. Such a shear motion related to recurrent jets was previously reported by \citet{2015ApJ...815...71C}. In their case, the satellite spot moved toward the southwest direction at an average velocity of 290 m s$^{-1}$ during the jet series, slightly lower than our observations. As the satellite spot studied here moved, its area gradually decreased due to magnetic cancellation with the surrounding positive magnetic field, eventually leading to its disappearance. Recurrent jets typically result in the disappearance of satellite spots with minority-polarity magnetic field \citep[e.g.,][]{2023ApJ...945...96Y}. During the recurrent jetlet eruptions, magnetic reconnection occurred continuously at the jetlet base, with an average cancellation rate of 1.3$\times$10$^{18}$ Mx hr$^{-1}$. This rate is similar to that for coronal jets \citep[$\sim$0.6$\times$10$^{18}$ Mx hr$^{-1}$,][]{2018ApJ...853..189P} and jetlets \citep[$\sim$1.5$\times$10$^{18}$ Mx hr$^{-1}$,][]{2018ApJ...868L..27P} in coronal holes, but lower than that for coronal jets in active regions \citep[$\sim$1.5$\times$10$^{19}$ Mx hr$^{-1}$,][]{2017ApJ...844...28S}. The average cancellation rate appears to influence the scale of small-scale eruptions.

\item Due to the high resolution of TiO images, the evolutionary details during magnetic cancellation can be investigated. During this process, dark lanes --- initially separated from the satellite spot and small pores --- were observed to move toward nearby satellite spots, small pores, or dark lanes with opposite polarities and eventually disappear. Dark lanes represent regions where cooler, denser gas is sinking, and where magnetic field is concentrated \citep{1998ApJ...499..914S}. The separation of dark lanes from the satellite spot or small pores suggests that these features was in a state of gradual disintegration before disappearing entirely. The separated dark lanes might play a key role in magnetic cancellation. It is notable that the satellite spot remained consistently redshifted throughout its evolution. \citet{1987ARA&A..25...83Z} proposed two models for magnetic cancellation: one is the $\textquoteleft$$\Omega$-loop submergence$\textquoteright$ model, and the other is the $\textquoteleft$U-loop mergence$\textquoteright$ model. In the first model, the $\Omega$-loop gradually submerges back into the Sun's interior after magnetic reconnection occurs above the photosphere. As the loop sinks, the magnetic field weakens, leading to magnetic cancellation at the surface, where the cancellation region exhibits a redshift. In the second model, magnetic reconnection occurs below the photosphere, allowing the U-loop to gradually emerge into the Sun's photosphere. As the loop rises, it causes magnetic cancellation at the surface, leading to a blueshift in the cancellation region. Clearly, our observations conform to the first model. Additionally, we note that the remaining satellite spot experienced a clockwise rotation and rapidly disappeared. We surmise that the weakening or reorganization of the magnetic field during the magnetic cancellation process may disrupt the original balance, thereby triggering the rotation of the satellite spot. The helicity of the satellite spot should be negative, which is consistent with that of its overlying flux rope. This suggests that the satellite spot continued to accumulate magnetic twist until it disappeared. 

\item It was found that the strongest jetlet (J4) was driven by the eruption of a micro-filament, which was clearly visible during its eruption. The length of the micro-filament was estimated to be approximately 6000 km, slightly exceeding the estimated range (500 km–5000 km) for jetlets associated with micro-filaments, as provided by \citet{2024ApJ...963....4S}. However, the scale of the micro-filament was based on an estimate derived from its base, rather than its actual length. Similar to the observations of \citet{2022ApJ...933...21K}, the micro-filament was only identified during its eruption. As mentioned earlier, the micro-filament might have been sheltered by large-scale loops, which could explain why it was invisible prior to its eruption. For the northward active region jets, \citet{2024ApJ...960..109S} suggested that their source region could be obscured along the line of sight by absorbing chromospheric material. Therefore, the invisibility of the micro-filament could be attributed to such visual effects. 

\item Both imaging and spectral observations reveal a counterclockwise rotation of J4, suggesting that the twist of the micro-filament was transferred to the jetlet through magnetic reconnection. In this event, the eruption of the micro-filament preceded 
that of the jetlet. The formation of the jetlet might be consistent with the two-step reconnection model for coronal jets proposed by \citet{2015Natur.523..437S}. Initially, internal reconnection took place within the expanded legs of the micro-filament's magnetic field, 
producing a bright point. This was followed by external reconnection (or null-point reconnection) occurred between the outer envelope of the erupting micro-filament field and its overlying fan-spine structure, leading to the 
formation of a jetlet along the outer spine of the fan-spine structure and the appearance of the circular base brightening. Rotational motion is a well-documented feature of coronal jets \citep[e.g.,][]{2013RAA....13..253H,2018MNRAS.476.1286J,2019ApJ...887..154L,2024MNRAS.530..473X}, but it is rarely observed in jetlets. According to \citet{2018ApJ...868L..27P}, only one out of ten jetlets has been reported to exhibit rotation. 
In this event, only J4 showed clear evidence of rotation. Particularly, the counterclockwise rotation of J4 is consistent with the helicity of the extrapolated flux rope prior to its eruption. Additionally, the horizontal magnetic field of the satellite spot consistently exhibited a vortex structure throughout its evolution until it vanished, indicating a significant distortion in the magnetic field configuration. The NLFFF extrapolation confirms that the satellite spot serves as one footpoint of a mini-flux-rope. Based on these observations, we propose that a mini-flux-rope or multiple mini-flux-ropes, repeatedly erupt along the magnetic neutral line as a result of continuous magnetic cancellation, giving rise to the recurrent jetlets observed. This idea is supported by previous studies. As analogues of coronal jets, jetlets might share similar physical mechanisms with coronal jets. Three-dimensional simulations of coronal jets by \citet{2016ApJ...827....4W} show that the repeated formation and ejection of flux ropes can result in intermittent jets. Furthermore, earlier observations demonstrate that magnetic cancellation is capable of triggering the formation and subsequent eruption of mini-filaments \citep{2011ApJ...738L..20H,2020ApJ...902....8C}. 

\end{enumerate}

In summary, the recurrent jetlets observed near the active region share certain physical characteristics with those of coronal holes, although they exhibit higher projected speeds. These jetlets are closely associated with a satellite spot moving northwest, whose magnetic field cancels with the opposite polarity magnetic field in the surrounding area, at a rate similar to that observed in coronal holes. During this cancellation process, several detailed features were observed, including dark lanes separated from the satellite spot and small pores, moved toward regions of opposite polarity and eventually disappeared. Additionally, one twisted jetlet was observed to be driven by a micro-filament eruption. The horizontal magnetic field in the satellite spot consistently exhibited a vortex structure throughout its entire evolution. These observations suggest that the recurrent jetlets are likely driven by three-dimensional null-point magnetic reconnection, triggered by the continuous eruption of a mini-flux-rope or multiple mini-flux-ropes, resulting from ongoing magnetic cancellation. This study presents compelling observational evidence that links the formation and eruption of mini-flux-ropes to recurrent jetlet activity. The findings align well with previous theoretical models and numerical simulations of coronal jets, implying that analogous magnetic processes operate at smaller spatial scales. Therefore, this work advances our understanding of the role of localized magnetic structures in solar atmospheric dynamics and provides new insights into the fundamental mechanisms driving small-scale solar eruptions.

\acknowledgments

The authors would like to thank the referee for their valuable feedback, which helped improve the manuscript. We thank Professor Hui Tian for his guidance. We are also grateful to the science teams of NVST, IRIS, 
and SDO for providing the data. This work is supported by the Strategic Priority Research Program of the Chinese Academy of Sciences, Grant No. XDB0560000, National Natural Science of China (12325303, 12433010, 12273108, 12173084,12273060, 12473059, 12203097, and 12303059), Yunnan Science Foundation of China (202301AT070349 and 202301AT070347), the Yunnan Science Foundation of China (202401AT070071), Youth 
Innovation Promotion Association, CAS (Nos 2023063), the Open Research Program of Yunnan Key Laboratory of Solar Physics and Space Science (YNSPCC202211 and YNSPCC202212), Young Elite Scientists Sponsorship Program by 
Yunnan Association for Science and Technology, Yunnan Revitalization Talent Support Program-Science $\&$ Technology Champion Project (No. 202105AB160001), Yunnan Provincial Department of Education Science Research 
Fund Project (No. 2025J0945),  and Chuxiong Normal University Doctoral Research Initiation Fund Project (No. BSQD2420).

\appendix

Four supplemental animations corresponding to Figures 2, 4, 7, and 8 are provided. These animations respectively depict the evolution of the recurrent jetlets, the J4 event, the satellite spot and nearby small pores with positive magnetic field at the base of the jetlets along with the associated magnetic flux evolution, and the satellite spot in a close-up view.

\clearpage

\begin{figure}
\centering
\includegraphics[scale=1.0]{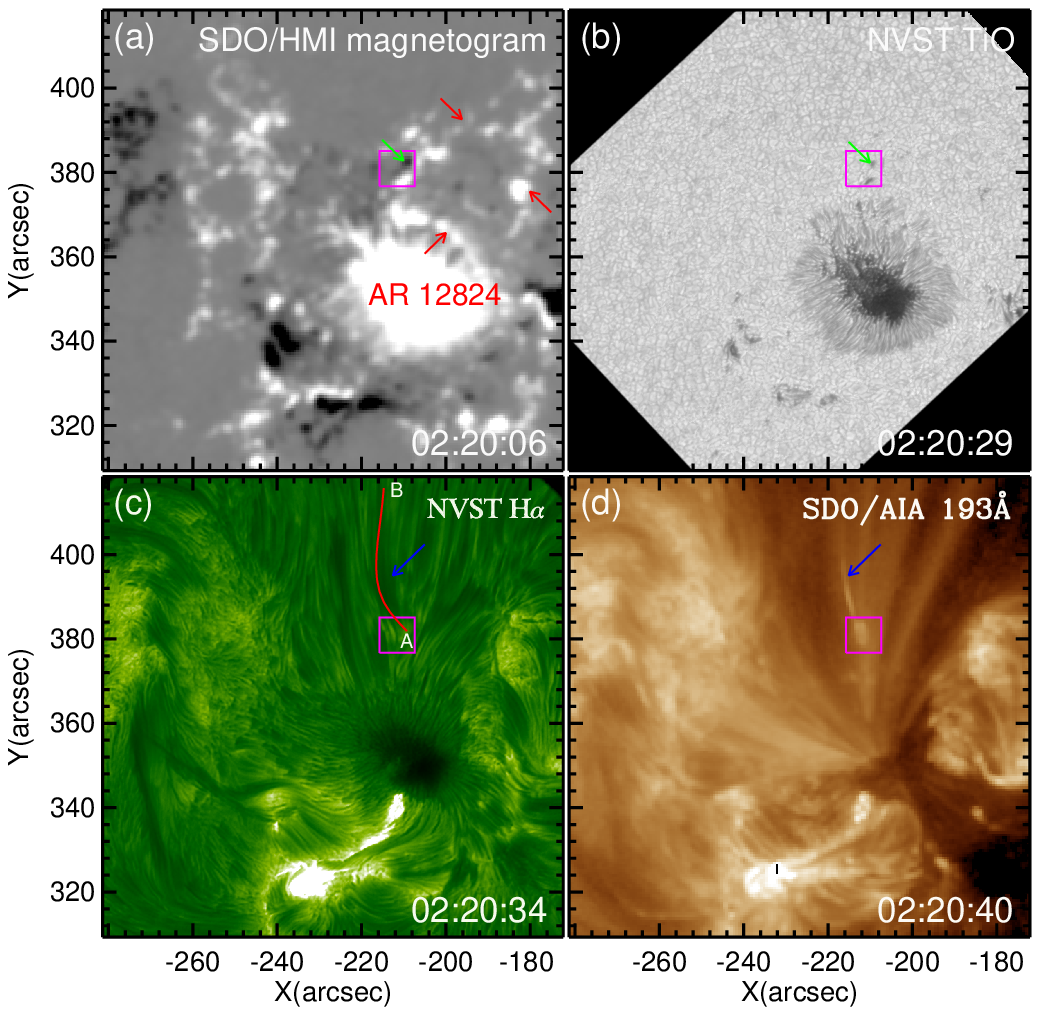}
\caption{Appearance of the active region that produces the recurrent jetlets. Panels (a) and (b) display the HMI line-of-sight magnetogram and NVST TiO image, respectively, marking the location of the recurrent jetlets. Panels (c) and (d) show the 
first jetlet in NVST H$\alpha$ and SDO/AIA 193~{\AA} images, respectively. The purple squares highlight the source region of the jetlets. The two green arrows in panels (a) and (b) mark the satellite spot, while the three arrows in panel (a) point to the positive magnetic network. The blue arrows in panels (c) and (d) indicate the first jetlet. \label{fig:fig1}}
\end{figure}

\clearpage

\begin{figure}
\centering
\includegraphics[scale=1.0]{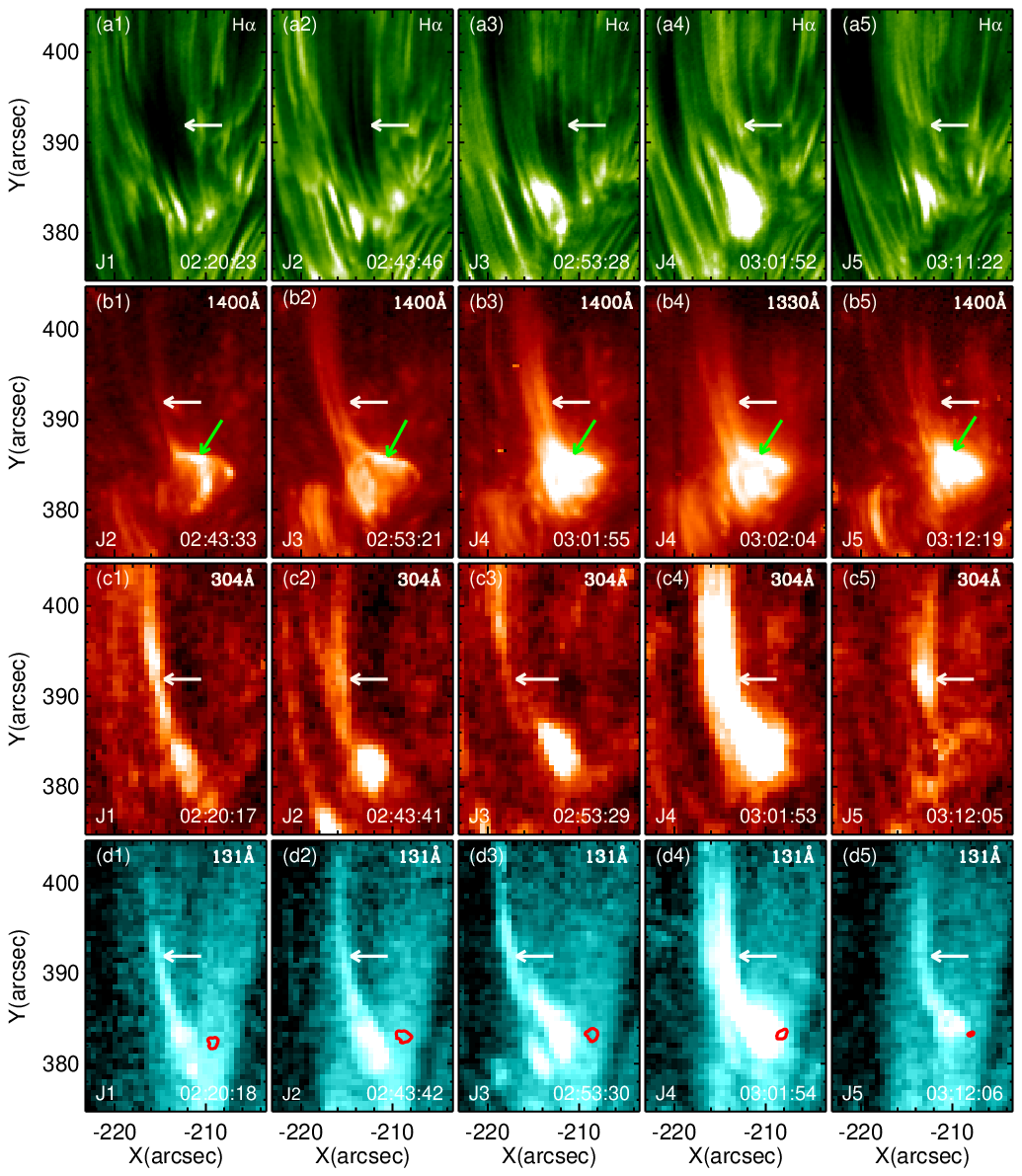}
\caption{The recurrent jetlets shown in different wavelengths. Panels (a1) -- (a5), (b1) -- (b5), (c1) -- (c5), and (d1) -- (d5) display the five jetlets in NVST H$\alpha$, IRIS 1400~{\AA} and 1300~{\AA}, SDO/AIA 304~{\AA}, 
and SDO/AIA 131~{\AA} images, respectively. The white arrows in all the panels point to the jetlets, and the green arrows in panels (b1) -- (b5) indicate the inverted-Y-shaped structures at the jetlet base. The red contours 
in panels (d1) -- (d5) depict the negative-polarity satellite spot from almost simultaneous NVST Tio images. An online animation of NVST H$\alpha$, IRIS 1400~{\AA}, SDO/AIA 304~{\AA} and 131~{\AA} wavelengths is available. 
The $\sim$4 s animation covers from $\sim$02:10 to $\sim$03:17 UT.  \label{fig:fig2}}
\end{figure}

\clearpage

\begin{figure}
\centering
\includegraphics[scale=1.0]{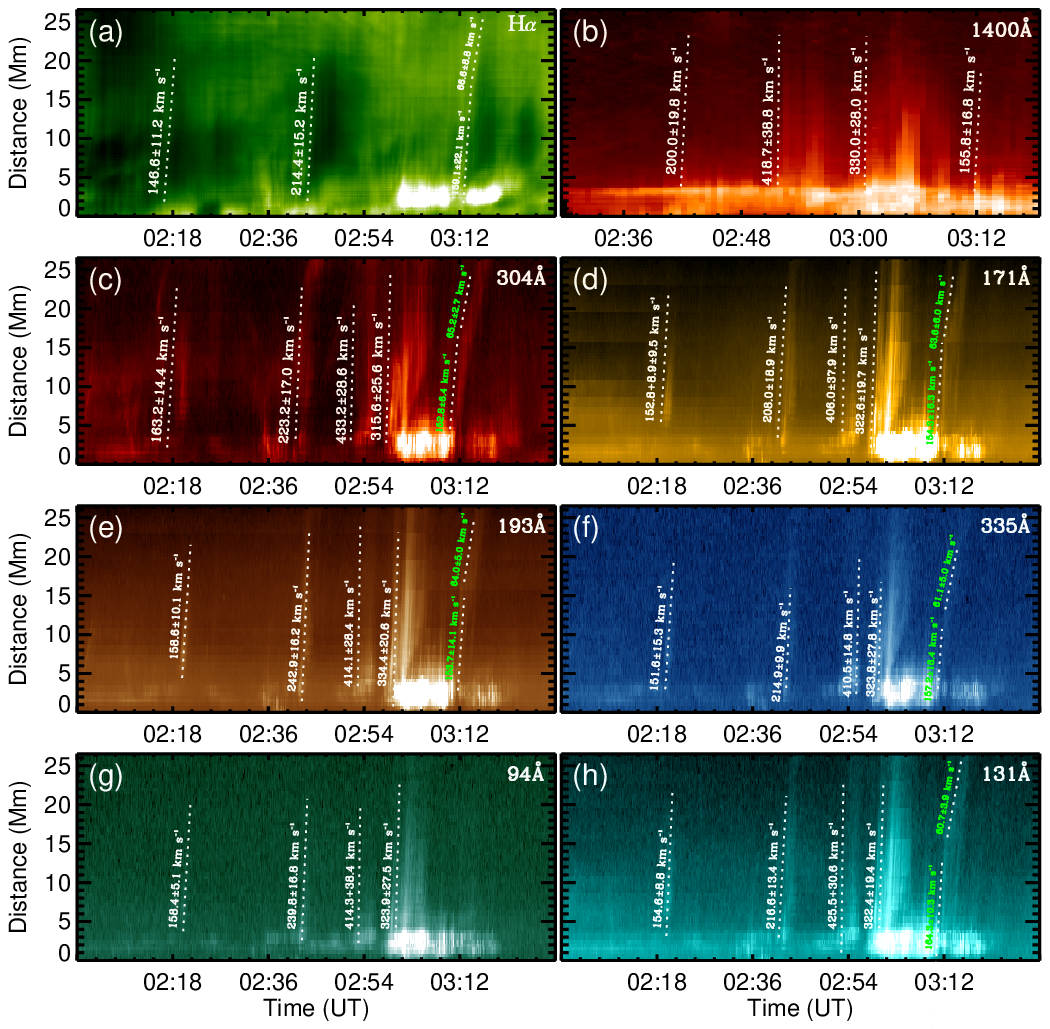}
\caption{The speeds of the jetlets. Time–distance diagrams along the line ``A'' -- ``B'', as marked in Figure 1(c), for H$\alpha$ (a), 1400~{\AA} (b), 304~{\AA} (c), 171~{\AA} (d), 193~{\AA} (e), 335~{\AA} (f), 94~{\AA} (g), 
and 131~{\AA} (h) wavelengths. The corresponding speeds are labeled on the diagrams. \label{fig:fig3}}
\end{figure}

\clearpage

\begin{deluxetable}{cccccc}
\tablecaption{The physical parameters of the jetlets\label{chartable}}
\tablewidth{700pt}
\tabletypesize{\scriptsize}
\tablehead{
\colhead{Number} & \colhead{Time (UT)\tablenotemark{a}} &
\colhead{Lifetime (minutes)} &
\colhead{Speed (km s$^{-1}$)\tablenotemark{b}} &
\colhead{Spire Width (km)\tablenotemark{c}} & Jetlet base (km)\tablenotemark{d}}
\startdata
  J1     &    02:18   &  3  & 152.8 & 1500 & 1300 \\
  J2     &    02:42   &  3  & 208.0 &  1300 & 2200 \\
  J3     &    02:54   &  3   & 406.0 & 1700 & 2400 \\
  J4     &    02:58   &  10   & 322.6 & 2900 & 2500 \\
  J5     &    03:11   &  3   & 154.3  & 1500 & 2600 \\
 \enddata
\tablenotetext{a}{The start time of the jetlet in SDO/AIA 171~{\AA} wavelength.}
\tablenotetext{b}{Projected speed along the jetlet spire measured from AIA 171~{\AA} wavelength.}
\tablenotetext{c}{Width measured in the middle of the spire using AIA 171~{\AA} images.}
\tablenotetext{d}{Cross-sectional width of the jetlet base during the jetlet onset in AIA 171~{\AA} wavelength.}
\end{deluxetable}

\clearpage

\begin{figure}
\centering
\includegraphics[scale=1.1]{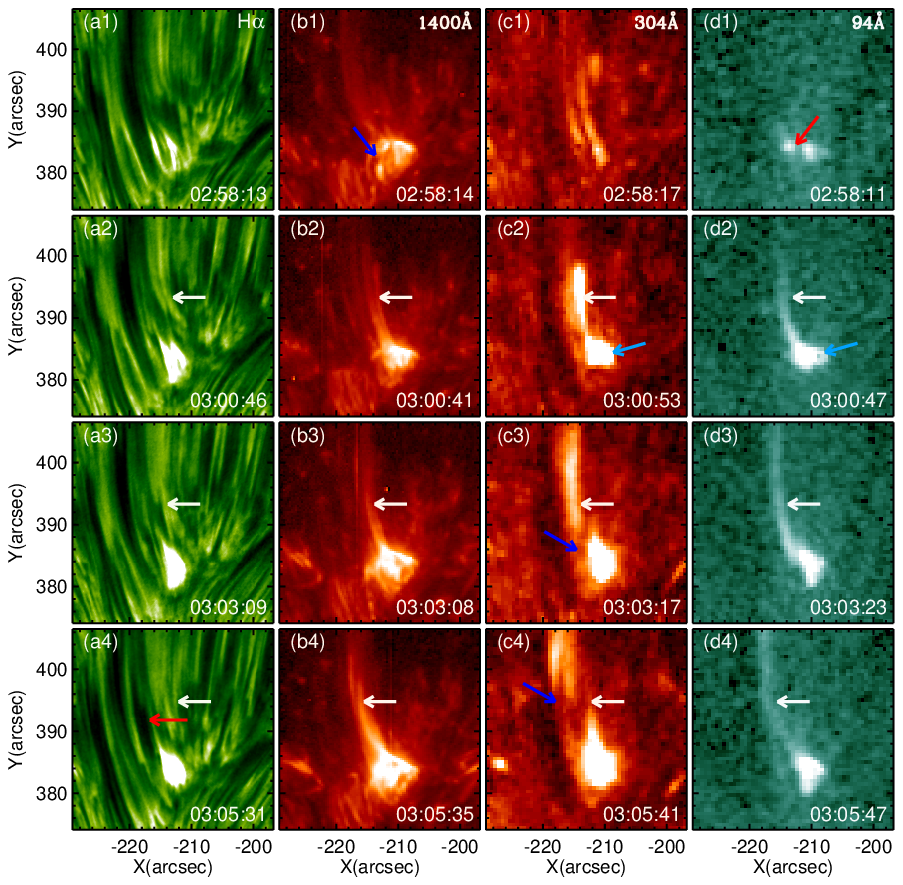}
\caption{The evolution of J4 across multiple wavelengths. Panels (a1) -- (a4), (b1) -- (b4), (c1) -- (c4), and (d1) -- (d4) display the evolution of J4 in NVST H$\alpha$, IRIS 1400~{\AA}, SDO/AIA 304~{\AA}, 
and SDO/AIA 94~{\AA} images, respectively. J4 is marked by white arrows in panels (a2) -- (a4), (b2) -- (b4), (c2) -- (c4), and (d2) -- (d4). The blue arrow in panel (b1) points to the macro-filament. 
The red arrow in panel (a4) and the blue arrows in panels (c3) and (c4) highlight the cool material of J4. The sky blue arrows in panels (c2) and (d2) mark the base brightenings. An online 
animation of NVST H$\alpha$, IRIS 1400~{\AA}, SDO/AIA 304~{\AA} and 94~{\AA} wavelengths is available. 
The $<$1 s animation covers from 02:57 to $\sim$03:09 UT. \label{fig:fig4}}
\end{figure}

\clearpage

\begin{figure}
\centering
\includegraphics[scale=1.0]{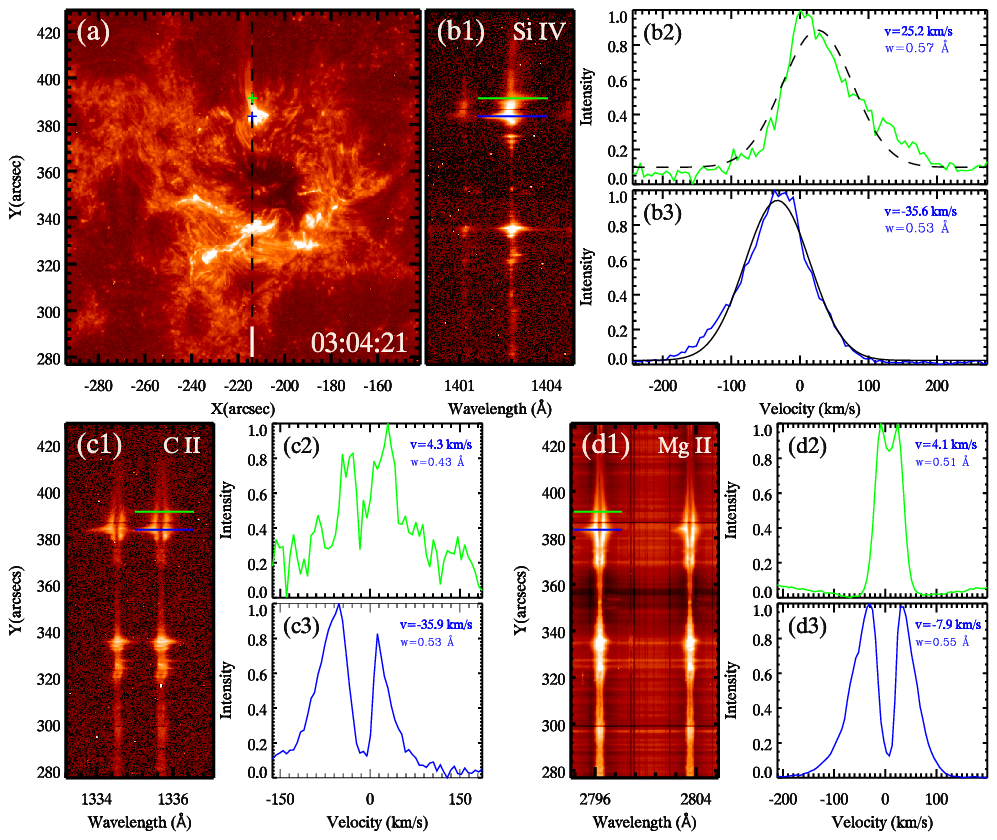}
\caption{J4 observed by IRIS. Panel (a) shows J4 in IRIS 1400~{\AA} slit-jaw image. Panels (b1) -- (d1) display IRIS spectra along the slit indicated by the black dashed line in panel (a). Panels (b2) -- (b3) present the profiles (green and blue solid
lines) and their one Gaussian fitting (dashed and solid black lines) of the Si $\tiny{\uppercase\expandafter{\romannumeral4}}$ 1402~{\AA} line at green and blue plus positions in panel (a). A quiet region along the slit, 
indicated by a white vertical bar in panel (a), was selected to obtain an averaged spectrum for determining the line center. Panels (c2) -- (c3) illustrate the profiles (green and blue solid lines) of the C $\tiny{\uppercase\expandafter{\romannumeral2}}$ 
1335~{\AA} line at green and blue plus positions in panel (a). Panels (d2) -- (d3) are similar to panels (c2) and (c3), but for the Mg $\tiny{\uppercase\expandafter{\romannumeral2}}$ k 2796~{\AA} line. All profiles are normalized to unity. \label{fig:fig5}}
\end{figure}

\clearpage

\begin{figure}
\centering
\includegraphics[scale=1.0]{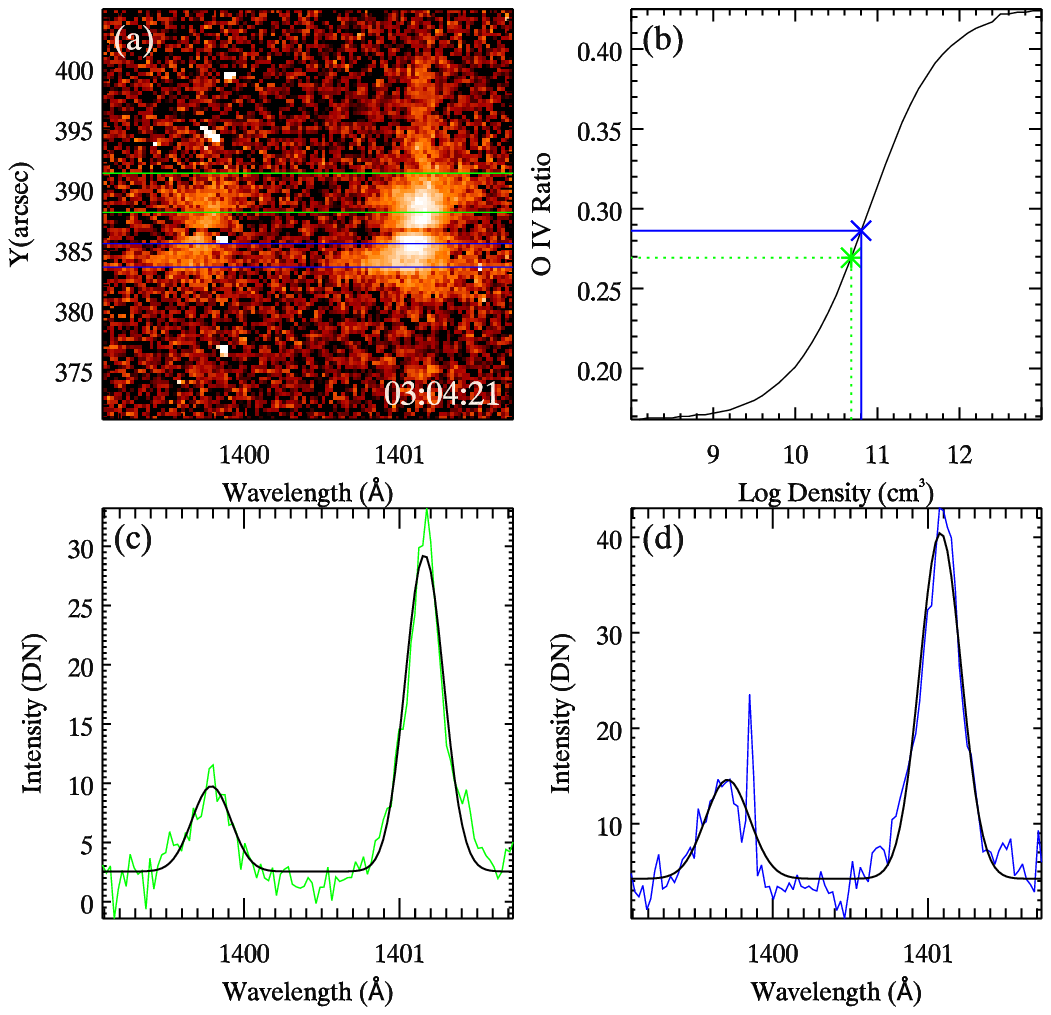}
\caption{The density of J4 and its base. Panel (a) shows the spectra of the two O $\tiny{\uppercase\expandafter{\romannumeral4}}$ lines along the slit indicated by the black dashed line in Figure 5(a). 
A specific position in the jetlet region/the bright point is selected, marked by two green/blue transverse lines. Panel (b) displays the theoretical ratios of the intensities of the two O $\tiny{\uppercase\expandafter{\romannumeral4}}$ lines 
versus the electron number density calculated using the CHIANTI atomic database v8.0. The colored lines represent the observed ratio of intensities of the O $\tiny{\uppercase\expandafter{\romannumeral4}}$ lines and 
electron densities. Panels (c) -- (d) present the observed averaged O IV profiles between two transverse green (blue) lines in panel (a) (the green and blue solid lines), along with the fitting results with double-Gaussian 
functions (the black curves). \label{fig:fig6}}
\end{figure}

\clearpage

\begin{figure}
\centering
\includegraphics[scale=1.0]{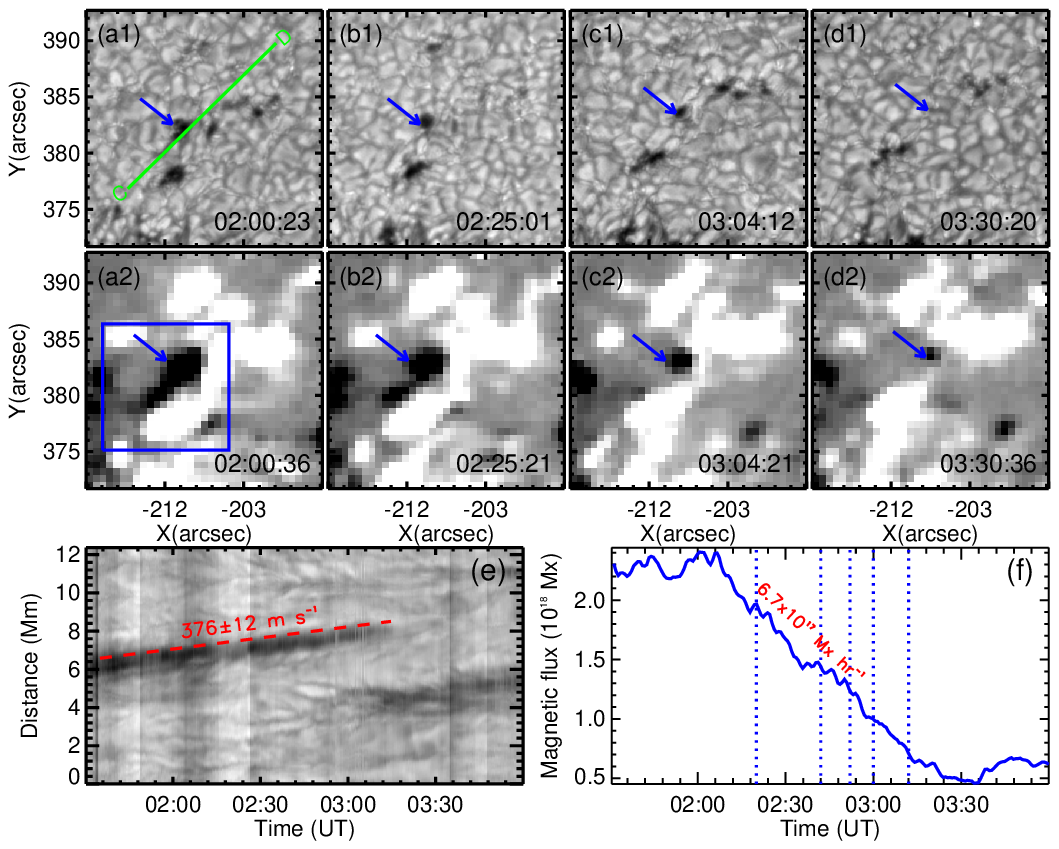}
\caption{Evolution of the satellite spot and nearby small pores with positive magnetic field at the jetlet base and the corresponding magnetic flux evolution. Panels (a1) -- (d1) show the evolution of the satellite spot 
at the base of the jetlets in NVST TiO images. Panels (a2) -- (d2) display the evolution of the corresponding magnetic field in SDO/HMI line-of-sight magnetograms. Panel (e) presents the time–distance diagrams along the line ``C'' -- ``D'' marked in panel (a1). 
Panel (f) illustrates variations of the negative magnetic fluxes in the blue box region, as shown in panel (a2). An online animation of NVST TiO images and SDO/HMI line-of-sight magnetograms is available. 
The $\sim$4s animation covers from 02:00 to $\sim$03:29 UT. \label{fig:fig7}}
\end{figure}

\clearpage

\begin{figure}
\centering
\includegraphics[scale=1.1]{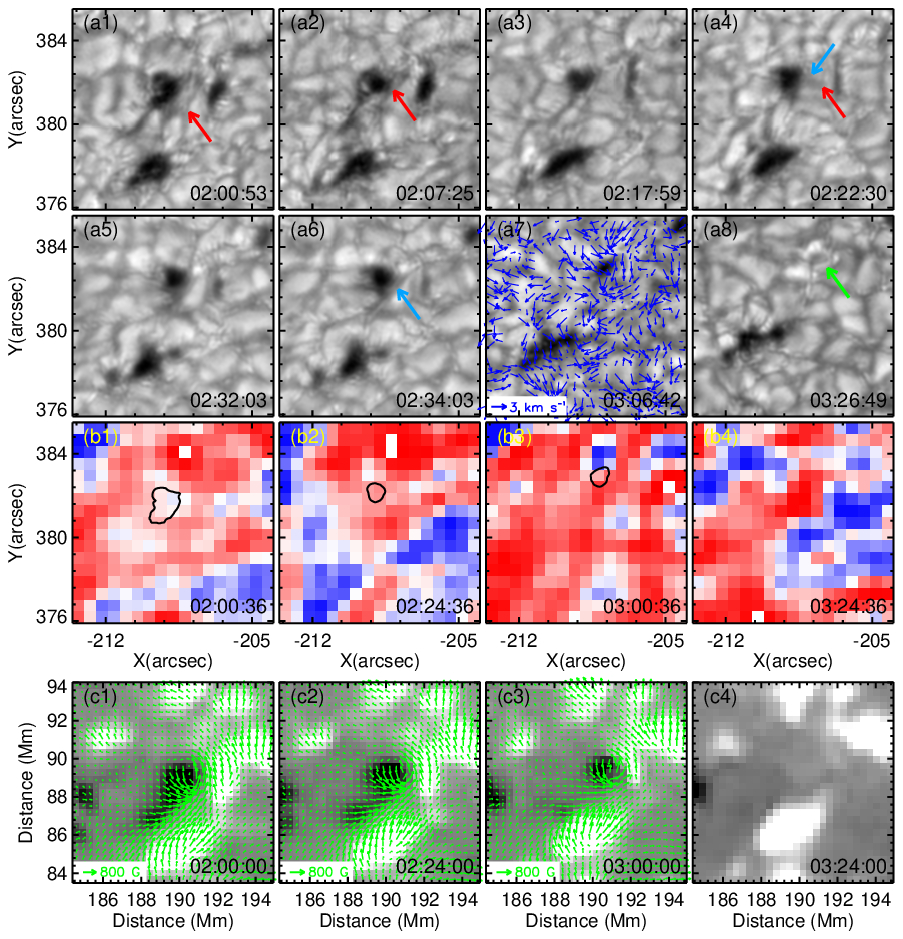}
\caption{The detailed evolution of the satellite spot and nearby small pores with positive magnetic field, along with their corresponding Doppler velocity and vector magnetic field at the jetlet base. 
Panels (a1) -- (a8) illustrate the evolution of the satellite spot at the jetlet base in NVST TiO images. 
 Panels (b1) -- (b4) show a sequence of SDO/HMI Doppler velocity images of the jetlet base, while panels (c1) -- (c4) display a sequence of SDO/HMI vertical magnetograms of the jetlet base. The blue arrows in panel (a6) are 
 the horizontal velocity field calculated using the dense optical flow method. The green arrows in panels (c1) -- (c3) are the horizontal magnetic field. The red arrows in panels (a1) -- (a3) point to the dark lanes separated from 
 the small pores with positive polarities, while the sky blue arrows in panels (a3) -- (a5) points to the dark lane separated from the satellite spot. The dark contours represent the satellite spot depicted from 
 the simultaneous TiO images. An online animation of NVST TiO images is available. The $\sim$7s animation covers from 02:00 to $\sim$03:29 UT. \label{fig:fig8}}
\end{figure}

\clearpage

\begin{figure}
\centering
\includegraphics[scale=1.0]{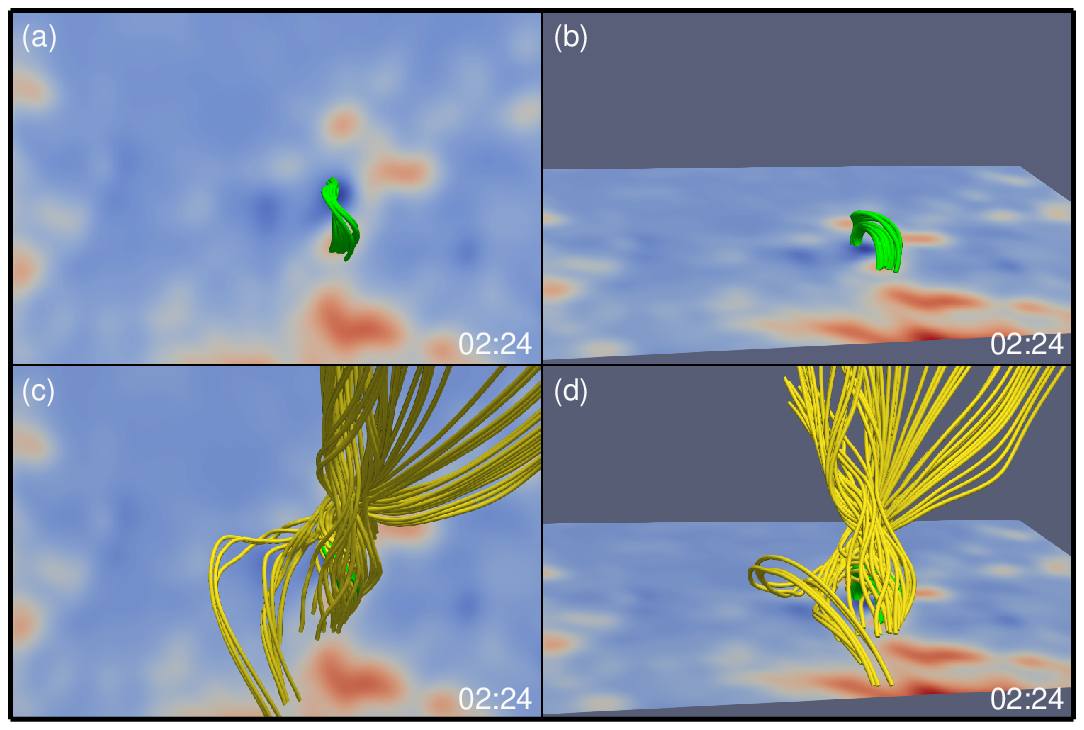}
\caption{The magnetic topology of the micro-filament and its overlying fan-spine structure, as revealed by the NLFFF extrapolation. Panels (a) and (c) show the top views of the extrapolated magnetic topology, while 
panels (b) and (d) show the side views. Red patches represent positive magnetic field, and blue patches indicate negative magnetic field. The green lines depict the micro-filament lying along 
the magnetic neutral line, while the yellow lines represent the fan-spine structure. \label{fig:fig9}}
\end{figure}

\end{document}